EIDGENÖSSISCHE TECHNISCHE HOCHSCHULE LAUSANNE
POLITECNICO FEDERALE DI LOSANNA
SWISS FEDERAL INSTITUTE OF TECHNOLOGY LAUSANNE

**COMMUNICATION SYSTEMS DIVISION (SSC)**
CH-1015 LAUSANNE, SWITZERLAND
http://sscwww.epfl.ch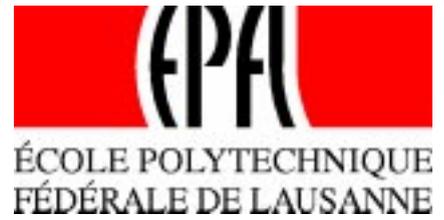

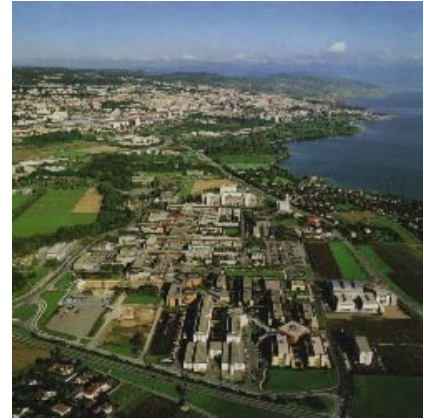

# Push vs. Pull in Web-Based Network Management

Jean-Philippe Martin-FlatinVersion 1: July 1998
Version 2: October 1998

Technical Report SSC/1998/022

# Push vs. Pull in Web-Based Network Management

**Jean-Philippe Martin-Flatin**
EPFL-ICA, 1015 Lausanne, Switzerland
Email: martin-flatin@epfl.ch   Fax: +41-21-693-6610   Web: http://icawww.epfl.ch

**Abstract**

In this paper, we show how Web technologies can be used effectively to (i) address some of the deficiencies of traditional IP network management platforms, and (ii) render these expensive platforms redundant. We build on the concept of *embedded management application*, proposed by Wellens and Auerbach, and present two models of network management application designs that rely on Web technologies. First, the *pull model* is based on the request/response paradigm. It is typically used to perform data polling. Several commercial management platforms already use Web technologies that rely on this model to provide for ad hoc management; we demonstrate how to extend this to regular management. Second, the *push model* is a novel approach which relies on the publish/subscribe/distribute paradigm. It is better suited to regular management than the pull model, and allows administrators to conserve network bandwidth as well as CPU time on the management station. It can be seen as a generalization of the paradigm commonly used for notification delivery. Finally, we introduce the concept of the *collapsed network management platform*, where these two models coexist.

## 1. Introduction

If we consider the design of an IP network management application with a software engineering perspective, it is a fairly simple case of distributed application. There are no stringent requirements put on it, such as real-time constraints or fault tolerance, and some management data may even be lost. Its complexity stems from only two points: there is a large, sometimes very large number of nodes to manage; and all management data traffic is considered as network overhead, and should therefore be kept to a minimum.

In the same perspective, if we analyze how IP networks are typically managed today, (that is, how network management platforms are designed, how efficient is SNMP as an access protocol, and how efficient is the principle of data polling inherent to the manager/agent paradigm), it is clear that most network management applications do not withstand the comparison with modern distributed applications. Why not use object-oriented analysis, design and implementation, which are widely adopted by the industry today? Why be limited by the few existing SNMP protocol primitives to collect data from an agent? Why incur the network overhead of having the manager repeatedly tell every agent what selection of MIB variables it is interested in, when this selection remains constant over time? Why not compress data efficiently when it is transferred between agents and managers? Why use an unreliable transport protocol to send a critical alarm to a management station when an interface goes down on a backbone router? Why make it so difficult to cross firewalls to manage remote subsidiaries? Why are management data transfers so often insecure?

In light of the technologies widely used today, many design decisions in IP network management appear inefficient or outdated. But they did not in 1988-90, when the first SNMP framework was devised. Moreover, if we place ourselves in a historical perspective taking into account how the market evolved [13], many deficiencies in today's commercial network management platforms can be analyzed and understood. The success of SNMP-based network management is due to a large extent to its simplicity, so it would be unfair to criticize this simplicity afterwards. Still, the way IP networks are typically managed in practice evolved very little throughout the 1990s. If IP network management continues to evolve so slowly, it runs the risk of going from simple to simplistic. This could result in a plethora of alternatives being proposed by multiple vendors, and in the end of open integrated network management.

As we showed in earlier work [12], there are several alternatives to traditional SNMP-based management: we have Web-based management, mobile agents, active networks, CORBA, intelligent agents, etc. In our view, Web technologies are the best candidate for improving this situation in the short term. The reason for this is fivefold. First, the solutions we describe in this paper are simple, and could be engineered and widely deployed in less than a year; mobile agents, conversely, require secure environments (especially for WAN links) which no one can provide currently; similarly, simple, yet efficient multi-agent systems for IP network management still remain to be seen. Second, Web technologies have a limited footprint on network devices, unlike CORBA. Third, not only do Web technologies bring solutions to the above-mentioned problems, as we demonstrate in this paper, but they also offer a smooth migration path, a key feature if they are to be adopted by the industry. Fourth, they allow keeping a coherent single framework for open network management, unlike WBEM. Fifth and last, the World-Wide Web has encountered lately a tremendous success in the enterprise world. Its simplicity, together with the portability of Java, have made it so ubiquitous that it is difficult today to find any software engineering field that is not using (or migrating to use) one of its early technologies (Web browsers, HTTP, HTML and CGI scripts) or one of its newer technologies (Java applications, applets, servlets, RMI and JDBC). Web expertise is rapidly developing worldwide, and it makes sense to capitalize on this in network management.

The idea of using the Web in IP network management is not new. Experiments with the early Web technologies (that is, Web browsers, HTTP, HTML and CGI scripts) started in 1993-94. Initially, they were only confined to secondary tasks. For instance, people developed HTML forms to standardize and automate problem reporting, which facilitated the work of calldesks. Network administrators also replaced daily, weekly and monthly printed reports with electronic versions put on an internal Web server. More interestingly, administrators began writing symptom-driven HTML forms that operators could use for routine network troubleshooting; the interactive interfaces provided by the Web proved to be much more user-friendly than the thick binders full of procedures that operators were used to. When network equipment documentations were shipped in electronic format, they were put on internal Web servers; not only were they easier to access, but administrators could then directly embed pointers to relevant pages of the documentation within symptom-driven HTML pages. This integration of documentation, procedures and tools was a step forward in network troubleshooting.

The first important step toward Web-based network management was taken when vendors began embedding HTTP servers in their network equipment. Bruins [2] reports some early experiments made by Cisco in 1995, whereby the entire command line interface was mapped to URLs. For instance, a Web browser could request <URL:http://router_name/exec/show/interface/ethernet0/> to a router, which would treat it as if the command line `show interface ethernet0` had been typed in interactively. This opened new doors for configuration management and symptom-driven HTML forms, as there was no more need to `telnet` into network devices. Mullaney [16] also describes work conducted by FTP Software, whereby agents send a static, locally stored, or a dynamically generated HTML page back to the management station in response to an HTTP `get` or `post` request.

The second important step was taken when Java applets appeared in Netscape's famous Web browser, in 1995. To the best of our knowledge, the new horizons that this technology opened up in network management were first published and advertised in the July 1996 issue of The Simple Times. The founding article by Wellens and Auerbach [25] introduced the concept of *embedded management application*, and showed the advantages of using HTTP rather than SNMP to vehicle data between managers and agents. Although the authors do not explicitly refer to applets in their article, the solution they propose is to transform an add-on (that has to be ported to many different management platforms and operating systems) into a single applet that can run everywhere. This applet is stored in the managed device, and uploaded by the administrator via a Web browser. Communication between the applet and its origin agent later relies on HTTP instead of SNMP. Bruins [2] explicitly refers to applets in his description of prototype work by Cisco; but in the scenario he describes, once the applet is uploaded, subsequent communication with the agent relies on SNMP, not HTTP, which is a poor use of applets as we will show in section 3.2.2.

Wellens and Auerbach's applet-based approach has now been adopted by many network equipment vendors, who embed HTTP servers and management applets in their equipment, but also by some network management platform vendors, who support Web browsers as front-ends to their network management platform.

Since the time of this proposal, many new technologies have appeared on the Web. Today, besides applets and Java applications, we can also use servlets, RMI, etc. All these technologies open new possibilities and enable new designs of network management applications. Leveraging on these new technologies, we propose to push Wellens and Auerbach's idea two steps further. First, we show that the design paradigm they propose is just one instance of a more general paradigm, the *pull model*, which can not only be applied to ad hoc management, like they do, but also to regular management. Second, we introduce a novel design called the *push model*. Unlike the pull model, it is not based on the request/response paradigm, but on the publish/subscribe/distribute paradigm. With this scheme, management data transfers are always initiated by the agent, like SNMP notifications delivery in pre-Web network management. The push model reduces network overhead, and moves part of the CPU burden from managers to agents.

The remainder of this paper is organized as follows. In section 2, we present a summary of the main shortcomings of traditional SNMP-based network management, and outline how Web technologies can address them. In sections 3 and 4, we present the engineering details of the pull model and the push model, and analyze the pros and cons of three communication technologies: HTTP, sockets and RMI. Finally, we introduce the concept of *collapsed network management platform* in section 5, and conclude with some perspectives for future work.

## 2. Problems with Traditional SNMP-Based Network Management

This section presents an overview of the problems encountered in traditional SNMP-based network management (that is, IP network management before the Web days), and describes how Web technologies can address them. These problems can be grouped into four categories: network management platforms, protocol efficiency, security, and transport protocol. The terminology used in this paper, as well as the model of a network management platform on which our analysis is based, are both presented in detail in [13].

### 2.1. Network management platforms

In a recent paper [13], we presented a brief history of IP network management before the Web days, showing how people came to use vendor-specific management GUIs (called *add-ons* when they are integrated in network management platforms). This paper also details the shortcomings of IP network management before the Web. To summarize, customers have four grievances: (i) network management platforms are too expensive, in terms of hardware and software; there should not be a need for dedicated hardware to manage networks; (ii) there should be unlimited support for third-party RDBMSs; today, customers are limited by the peer-to-peer agreements that have been signed, or not signed, between RDBMS vendors and network management platform vendors; if they want the latter to support another RDBMS that they happen to own already, they are charged enormous amounts of money for the "port"; (iii) for the sole purpose of network management[1], some customers must support a Unix system, although they run a business entirely based on PCs and/or Mac's; they want to use a PC or a Mac instead, but they do not want to buy a whole new (and expensive) network management platform.

The answer of Web-based network management to grievance (i) is the collapsed network management platform, that we will gradually introduce in this paper. Grievance (ii) is addressed by JDBC, although there is a problem with regards to the poor execution speed of Java interpreted bytecode (even when speed-up techniques are used, such as the JIT compiler). Grievance (iii) can be solved by the platform independence of Java and the universal interface offered by Web browsers.

Network equipment vendors, on the other hand, are dissatisfied primarily by the huge costs they have to bear to support device-specific management GUIs for their equipment. To customers, a given GUI looks more or less the same, no matter what management platform is used underneath. But to network equipment vendors, it does not. When a new management GUI is released, the code has to be ported to many different operating systems (Windows 95, Windows 98, Windows NT 4.x, Solaris 2.x, HP-UX 10.x, HP-UX 11.x...) and many different management platforms supporting different APIs (HP OpenView, Cabletron Spectrum, Sun Solstice, IBM NetView...). Over time, despite the relatively small number and the stability of the major management platform vendors, the number of devices supported by each vendor and the number of operating systems to port to have grown so large that the maintenance costs of these management GUIs have skyrocketed.

With Web technologies, this problem is solved by applets, as we will see in section 3: the multiple incarnations of the same add-on are all replaced with a single piece of code, the management applet, written in Java.

Customers and network equipment vendors share two other concerns. First, they both want the time-to-market of management GUIs to be reduced. When they purchase a brand new piece of equipment, customers want to be able to manage it immediately via their favorite management platform. But many months can pass between the time a new network device that has been trumpeted by marketing is finally released and sold to customers, and the time its vendor-specific management GUI has been ported to all operating systems and all existing network management platforms. There are many environments where the constant availability of the network is critical to the smooth running of the business, and network equipment cannot be purchased unless it can be managed. So, for large companies that have peer-to-peer agreements with all major management platform vendors, there is a time window during which they cannot sell to these customers; this is a problem for customers and vendors alike. For small companies, and especially for start-up companies specialized in cutting-edge technology, this problem is even worse. As their market share is close to zero, they are of no interest to management platform vendors, who do not bother signing peer-to-peer agreements with them. Hence, customers who want to buy from small companies are reduced to managing their network equipment with either user-unfriendly MIB browsers, or with tailor-made software running on a dedicated PC sitting next to the device. Consequently, many markets are closed to such start-ups, which are desperate to get access to integrated network management.

---

1. Until roughly 1995, Windows-based network management platforms were not powerful enough to manage large networks and run large RDBMSs: in such environments, you had to buy a Unix system. Since then, the power of PCs has increased dramatically, much more than the power of Unix workstations. Customers who buy a management platform today are not exposed to this problem anymore.

The second problem, which concerns customers and vendors alike, is versioning [16]. When upgrading a vendor-specific MIB and consequently a vendor-specific management GUI, customers and network equipment vendors want to cope with situations where the add-on integrated to the management platform has a different version level from that of the MIB supported by the agent. Today, since there is no such a thing as a MIB-discovery protocol, administrators either have to manually specify what MIB is supported by what device, which is tedious, or they have to refrain themselves from using MIB variables that have changed between the last and the previous MIB versions, which can cause problems.

These last two concerns are again addressed by applets in Web-based management. Applet-based management software is embedded in a network device when you buy it, so you can manage your agent at once. When you upgrade the software on your agent, you can easily upgrade the management applet as well. And by transferring the management software from the agent to the manager, we ensure that the version of the vendor-specific MIB is always the same on both sides.

## 2.2. Protocol efficiency

Since the outset, SNMP-based network management has been hampered by two protocol engineering decisions which drastically reduce its efficiency. First, both SMIv1 [20] for the SNMPv1 framework, and SMIv2 [3] for the SNMPv2 and SNMPv3 frameworks, make the use of BER encoding [10] mandatory for SMI MIB data. Unfortunately, this encoding is renown for its inefficiency. Mitra [15] and Neufeld and Vuong [17] describe this issue in detail, and show that the amount of administrative data (identifier and length) transferred is very large compared to the actual data (content). Since ASN.1 itself does not mandate the use of any specific encoding rules, other more efficient schemes were defined, such as PER [11]. But they did not make their way through to the SNMP frameworks. The second issue is in SNMP itself. SNMP varbind lists are relatively expensive, because the OIDs used to name variables usually take much more space than the values. Also, the absence of an efficient table retrieval mechanism means that the total protocol efficiency suffers from repeated message exchanges (and repeated computations on the agent side).

These issues are addressed in Web-based network management by using HTTP 1.1 instead of SNMP to transfer SMI MIB data between managers and agents. The advantages are fourfold. First, this migration makes it possible to abandon BER encoding, and to use instead a new MIME content type for SNMP, or simply encode SMI MIB data in HTML [16]. Second, the use of persistent connections [6], a key feature of HTTP 1.1, alleviates the network overhead and latency induced by multiple TCP connection setups and teardowns. Third, pipelining [6], another key feature of HTTP 1.1, allows the manager to make multiple requests without waiting for each response. This reduces latency, but also allows a very efficient use of TCP connections, when combined with persistent connections: if the time-out value of each persistent connection is greater than the polling frequency for that agent, the same TCP connection can be used indefinitely between the manager and each agent. Fourth, the network bandwidth usage can be reduced by performing transparent data compression. Unlike SNMP, HTTP supports the MIME concepts of *content type* and *content transfer encoding* to transfer data. Therefore, it is possible to compress the payload of an HTTP packet (say with `gzip`) on the agent, and uncompress it on the manager, without the management application being even aware that data is compressed when it is in transit. Because the payload is plain text, the expected compression rate is fairly high.

The only problem not addressed by HTTP is the lack of an efficient table retrieval mechanism. This can be dealt with by adding a new primitive to the new MIME content type mentioned above. RMI and Object Serialization offer a neater solution, since they replace communication protocols like SNMP or HTTP with direct object-to-object communication. SNMP varbind lists are replaced with serialized objects, and the absence of an efficient table retrieval mechanism only affects the agent, as we will show further on. But there are also problems with RMI, as we will see in section 4.2.2.

## 2.3. Security

Security is a weak point of the SNMPv1 and SNMPv2 frameworks [22]. The lack of secure SNMP `get`'s and `set`'s has hampered the management of remote subsidiaries for many years. With SNMP, how can an enterprise reasonably manage a VPN spanning over the Internet or some kind of public network? Things have been significantly improved in this respect by the SNMPv3 framework, which was recently released. But field tests still remain to show that administrators are happy with this new security framework in deployment. Today, organizations that demand secure communications over public WAN links (e.g., banks) generally resort to expensive devices that perform transparent encryption and decryption at low protocol layers, at the boundaries of the WAN links. Whether the data is SNMP or else, it is always encrypted.

Thanks to SSL [23], HTTP has proved superior to SNMP, with respect to security, for several years. SSL is a security protocol which sits between the transport-layer protocol (TCP) and the application-layer protocol (HTTP), and prevents eavesdropping, tampering and message forgery. When a URL starts with (i.e., has an RFC-1738 scheme name equals to) `https`, the HTTP client connects to port `443/tcp` [9] on the server, instead of the standard HTTP port `80/tcp` [9]; then HTTP traffic is sent over SSL. Like in network management, much work is being done at IETF and W3C on Web security. This includes the successor of SSL, TLS [5], over which HTTP can be layered [18], but also the support for strongly secure electronic transactions, and the integration of strong authentication and privacy schemes in HTTP.

## 2.4. Transport protocol: TCP vs. UDP

The idea of using HTTP instead of SNMP is very intuitive: they are both request/response protocols, implemented with client/server technology, and an SNMP `get` or `set` can be mapped nicely to an HTTP `get` or `post`. But the fact that these communication protocols rely on different transport protocols immediately raises the question: Should management data be transported over a reliable protocol such as TCP, or an unreliable protocol like UDP? This issue has been debated for years. People often have strong opinions about it. Rose, one of the designers of SNMP, is fiercely opposed to using a reliable transport protocol [19]. Wellens and Auerbach, conversely, denounce what they call the *myth of the collapsing backbone* [25]. Their analysis takes into account the fact that in 1996, HTTP 1.0 was suffering from the lack of persistent connections, so the use of TCP connections for each SNMP `get` entailed a significant overhead and latency. As we know, HTTP 1.1 addresses that, which gives today even more power to their argument.

We believe that network administrators should have the choice. Not only should the selection of the transport protocol be customizable at the enterprise level, but it should also be at the network device level, and possibly even at the MIB variable level. Unfortunately, today, there is no such choice: everyone uses SNMP over UDP, although nothing prevents the use of SNMP over TCP. Based on our professional experience, we can state that there are settings where management data is more important than user data. In these cases, since management data traffic and user traffic share the same medium in the IP world, management data should either be allocated a certain (low) proportion of the link capacity (a feature already offered by some equipment vendors), or given a higher priority in routers (a feature hardly ever used today, but which may become common in the near future with the increase of real-time multimedia traffic). Clearly, it is the duty of the administrator to ensure that the proportion of the management data remains low compared to the link capacity, since the *raison d'être* of networks is to carry user data, not management data! Unlike WAN links, most intranets are oversized, and losses are rarely due to ongoing network saturation. They are rather due to mundane reasons such as Ethernet collisions, or temporary buffer overflow in a router in case of bursty traffic. Thus, in our view, in such networks, management data should be transported with a reliable protocol. Conversely, for backbone routers with impressive MTBFs, or for expensive intercontinental WAN links, management data is generally far less important than user traffic; in these cases, management data should be carried over an unreliable transport protocol such as UDP.

Above, when we said "management data", we meant data collection and network monitoring. SNMP notifications are a different case altogether. They are only sent in case of major problems. Because the definition of what is a major problem is site specific, most if not all network devices allow administrators to filter what notifications should or should not be sent to the NMS. Unfortunately, with the current SNMP technology, these notifications are also sent over an unreliable transport protocol. In our view, they should instead be given the highest possible priority. If an interface goes down in a backbone router, it is more important that the router reports this event back to the NMS immediately, than to actually route user traffic on the alternate path. Using UDP instead of TCP means that this important packet may be lost for a silly reason like a mere Ethernet collision on the NMS local network segment. Although TCP does not guarantee the delivery of packets, at least it supports automatic retransmissions and acknowledgments. UDP does not, and places the burden of retransmitting lost datagrams on the management application, making it unnecessarily complex.

In summary, data collection and network monitoring may use reliable or unreliable transport protocols, depending on the site-specific needs. TCP-based network management is probably a bad idea for heavily-used backbone routers, which are used flat out, need to work 24x24 and 7x7, and rarely experience any problems. But it seems well suited to intranets, especially for SMEs with small or medium-sized networks. As for notification delivery, TCP is probably better than UDP, because major problems really ought to be reported to the NMS over a reliable transport protocol, to maximize the chances of the notification arriving at destination.

An important side-effect of selecting UDP or TCP as a transport protocol is the issue of firewalls. More and more organizations protect their internal network from Internet intruders by setting up firewalls. In the commercial world, firewalls are expected to become ubiquitous in the near future. Most firewalls are setup to let HTTP traffic go through, because of the Web, and to filter out UDP traffic, because it can be the vehicle of well-known attacks [4]. This is a problem for companies that need to manage remote offices via WAN links. Some firewall systems support UDP relays, that dynamically learn about UDP traffic, and try to work out whether a measured pattern looks like an attack or normal usage. Such relays require ad hoc configuration of the firewall. Whether security or network administrators are willing to open up (even partially) external access to an internal NMS is primarily a matter of security policy, and is therefore site specific. But many may not want to, especially SMEs with remote offices that do not have in-house expertise regarding firewalls. For such companies, the use of TCP as a transport protocol may be the best option when they decide to protect themselves behind a firewall.

## 3. The Pull Model

In section 3.1, we first explain how the pull and push models work. In section 3.2, we then present the engineering details of pull-based ad hoc management, and show how Web technologies can complement the management functionalities offered by the NMS. In this case, network troubleshooting can be done from any machine running a Web browser, whereas the NMS remains in charge of regular management. Finally, in section 3.3, we show that Web technologies can also deal with regular management, and make the network management platform redundant for troubleshooting and data collection.

### 3.1. Pull vs. Push: the newspaper metaphor

In software engineering, the pull model and the push model designate two well-known approaches for exchanging data between two distant entities. The newspaper metaphor is a simple illustration of these models: if you want to read your favorite newspaper everyday, you can either go and buy it every morning, or subscribe to it once and then receive it automatically at home. The former is an example of pull, the latter of push. The pull model is based on the request/response paradigm (called *data polling,* or simply *polling,* in traditional SNMP-based network management); the client sends a request to the server, then the server answers, either synchronously or asynchronously. This is functionally equivalent to the client "pulling" the data off the server. In this approach, the data transfer is always initiated by the client, i.e. the manager. The push model, conversely, is based on the publish/subscribe/distribute paradigm. In this model, agents first advertise what MIBs they support, and what SNMP notifications they can send; the administrator then subscribes the manager (the NMS) to the data he/she is interested in, specifies how often the manager should receive this data, and disconnects. Later on, each agent individually takes the initiative to "push" data to the manager, either on a regular basis via a scheduler (e.g., for network monitoring) or asynchronously (e.g., to send SNMP notifications).

### 3.2. Ad Hoc Management

The simplest and most intuitive application of the applet technology is ad hoc management. In this section, we sum up the specificities of ad hoc and regular management, then come back to Wellens and Auerbach's model and study its engineering details; we show that generic GUIs can similarly be coded as applets, and conclude with a figure illustrating how ad hoc management can rely entirely on Web technologies.

#### 3.2.1. Ad hoc management vs. regular management

When they described the embedded management application approach, Wellens and Auerbach had ad hoc management in mind. Ad hoc management is always manual, and requires a user (administrator or operator) to interact with the management software via some GUIs. It is typical of transient tasks: you connect to a network device, retrieve some data to check something, and disconnect shortly after. Regular management, conversely, is concerned with ongoing data collection, network monitoring and event handling. It is automated to a large extent, and generally runs continuously.

Ad hoc management takes place in virtually all companies. In large organizations that can afford staff dedicated to monitoring the network (operators), or that rely on entirely automated regular management, ad hoc management is complementary to regular management. For instance, an administrator may occasionally want to pop up a GUI to configure a router, or an operator may take a look at a time series of error rates while investigating a network problem. Conversely, in smaller organizations such as SMEs, ad hoc management generally replaces regular management. There is no operator and no dedicated NMS: the management software is only used occasionally, on an ad hoc basis. Ad hoc management typically consists in troubleshooting (i.e., a network problem just showed up, and the administrator tries to identify and fix the problem manually), or configuration management (e.g., the administrator sets up a new router, or checks if a router is configured as expected).

### 3.2.2. Vendor-specific management GUIs coded as applets (embedded management applications)

As we saw in section 2, applets address many issues in SNMP-based network management. They decrease vendors' development costs for management GUIs; they address the issue of having different versions of a vendor-specific MIB in the same network; they cut the time-to-market of management GUIs down to zero; and they are independent of the machine where the Web browser runs: this can be a PC running Windows or Linux, a Mac, a Unix workstation, etc.

The uploading of the applet by the Web browser is depicted in Fig. 1. But how do we specify an applet for a given device? Like with traditional network management platforms, the entry point for the user (administrator or operator) is a map of the network, which is retrieved from an internal Web browser. This map can be a tailor-made applet, or more simply a GIF image used as an HTML sensitive map: when a user clicks on it to select the icon of a network device, the (x,y) coordinates are mapped to agents by a CGI script, and the corresponding URL is requested from the selected agent. The HTTP server running on the agent retrieves its vendor-specific management applet from local storage, e.g. from EPROM, and sends it back to the Web browser. Once the applet is uploaded by the Web browser, there are two ways to proceed: either use SNMP or HTTP.

If we use SNMP, the interactions between the manager (Web browser running on any machine) and the agent (network device) are depicted in Fig. 1. Steps 1 and 2 occur once, to transfer the applet, whereas steps 3 and 4 are an iterative process. The dotted arrow for step 2 is a visual aid that shows that the applet is transferred from the agent to the manager (as opposed to two entities simply communicating with each other). In reality, this transfer takes place between the HTTP client of the Web browser and the HTTP server of the agent. Once the applet is uploaded, the user has the equivalent of an add-on in an NMS to interact with. Graphical interactions are translated into SNMP commands by the applet (e.g., a mouse click on the drawing of a reset button can be mapped to an SNMP `set`). In other words, we have a Java API making SNMP calls underneath.

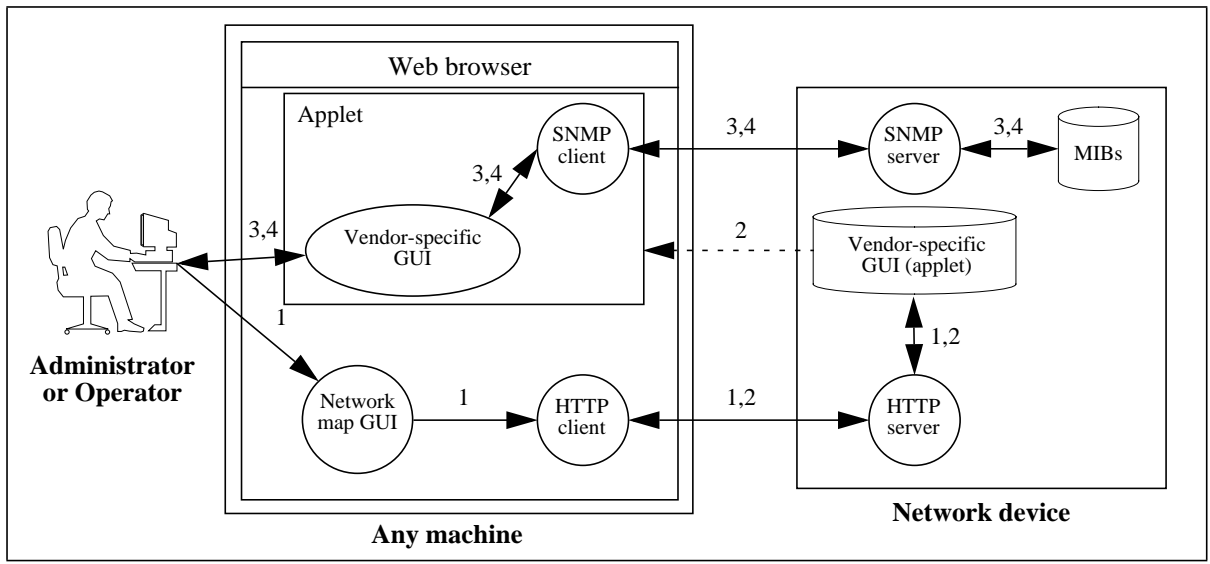

**Fig. 1.** Pull model: HTTP together with SNMP

But how can we load this SNMP stack in the browser? The simplest approach is to include an entire SNMP stack in the applet, as described by Bruins [2], because the applet security model prevents it from retrieving this stack from the local file system. So each time a management applet is uploaded from a network device, the entire SNMP stack needs to be moved along. This is clearly inefficient, especially if this transfer takes place across a WAN link. An improvement on this is to retrieve the SNMP stack separately, via a socket. Because of the applet security model, sockets may only be opened between an applet and its origin server; so we need a proxy to act as the origin server. The management applet is first requested by the Web browser to the proxy; second, the proxy contacts the network device and retrieves the applet without the SNMP stack; third, the applet is passed along to the Web browser which executes it; fourth, the applet opens a socket to the proxy, which has to run the server side of the socket (that is, an application like a Unix daemon); fifth, the SNMP stack is transferred via the socket and the socket is closed; after that, all SNMP traffic between the manager and the agent is simply relayed "as is" by the proxy.

This approach is of limited interest though, because by not using HTTP between the Web browser and the agent, we do not benefit from the advantages of HTTP over SNMP presented in section 2: improved protocol efficiency, by encoding SMI MIB data (MIB variables) in a specific MIME type, or embedding it in an HTML structured document; improved security, by using SSL, etc. So let us now study the case when the recurrent steps 3 and 4 are based on HTTP rather than SNMP. Fig. 2 shows the engineering details of this solution. This time, we no longer use SNMP between the manager and the agent; consequently, there is no need for an SNMP client in the Web browser, thus no need for a proxy.

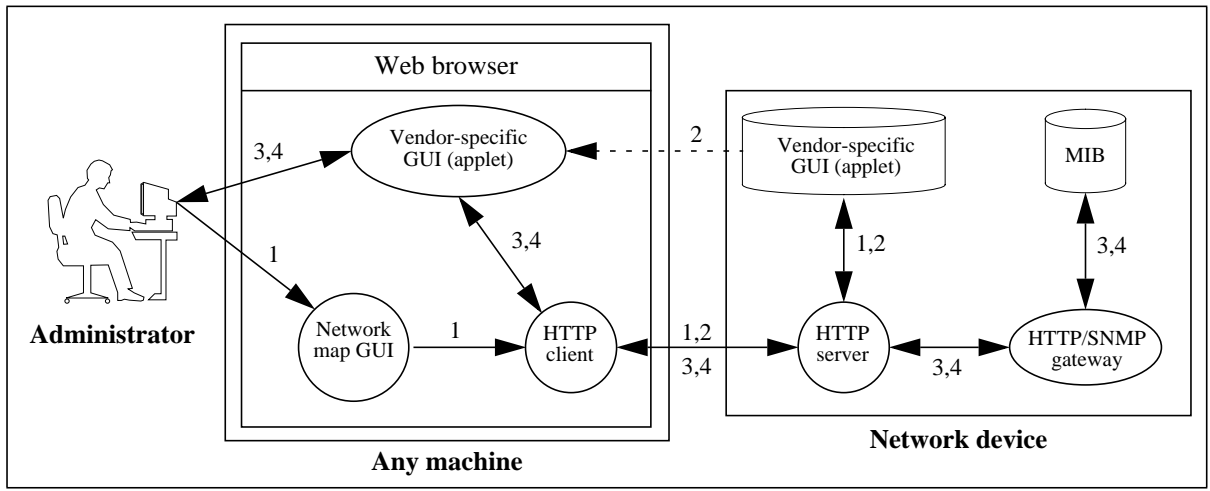

**Fig. 2.** Pull model: HTTP instead of SNMP

In Fig. 2, the "MIB" icon in the network device represents the vendor-specific MIB of the network device. This icon is represented as if the MIB were stored on local storage, like the management applet, because a MIB is a virtual data repository; in practice, of course, it is not stored on local storage, because that would be grossly inefficient, but as data structures in memory.

When a MIB variable is requested by the applet, the request is made to the HTTP server run by the agent. This HTTP server then launches an HTTP-to-SNMP gateway to access the local MIB. Depending on the degree of optimization of the code run by the agent, this gateway can either directly access the MIB data structures in memory, or do an explicit SNMP `get` or `set`. This gives a useful migration path to network equipment vendors.

Vendor-specific management applets are commercial in nature. Just like you have the possibility to buy add-ons and integrate them in pre-Web network management platforms, you can also buy a management applet when you purchase a network device. Depending on the pricing policy of your network equipment vendor, you may lose out if you already have several of the same equipment, because the overall cost of all applets (which are sold independently from each other) exceeds the price of the add-on (shared by all devices). You may also gain a lot if you have a very heterogeneous network. Since an applet is commercial software, it requires some kind of licence enforcement technology. Rather than check for a valid licence each time an applet is downloaded, which would require extra software in all network devices, but also the management of licence keys in the CGI script invoked by the HTML sensitive map, it may be easier for vendors to sell the HTTP server as an extra chip in their equipment, and make their applet freely available on their Web site. Licence enforcement would no longer be performed by software, based on the exchange of a key, but by hardware.

### 3.2.3. Generic management GUIs coded as applets

Up to now, we followed the original idea of Wellens and Auerbach: only vendor-specific management GUIs are coded as applets. But it is very reasonable to code generic GUIs[1] as applets, too. In this case, ad hoc management relies entirely on Web technologies: the SNMP-based network management platform need only cater to regular management.

Unlike vendor-specific management applets, generic management applets may either be available free of charge, or sold. For such widely used MIBs as MIB-II or RMON, one can very reasonably expect that free software of good quality can be retrieved from the Internet. AdventNet's SNMP package [1] and Sun's JMAPI [24] are two examples for MIB-II. Other more arcane MIBs such as the UPS MIB may require software sold by a network management software vendor, since they may not raise much interest in the Internet freeware development community.

---

1. Generic GUIs are management software supporting generic MIBs, e.g. MIB-II, the RMON MIB, the ATM MIB, the FDDI MIB, etc.

Fig. 3 depicts the technical aspects of this solution. The Web server can be any machine on the Internet or the intranet. The "MIBs" icon in the network device represents all the generic MIBs supported by the agent, plus its vendor-specific MIB. In this figure, we added a dash-dot line to epitomize the possible presence of a firewall between the manager and the agent. We will come back to firewalls further on, but let us point out that for ad hoc management, we just need to have HTTP traffic go across the firewall.

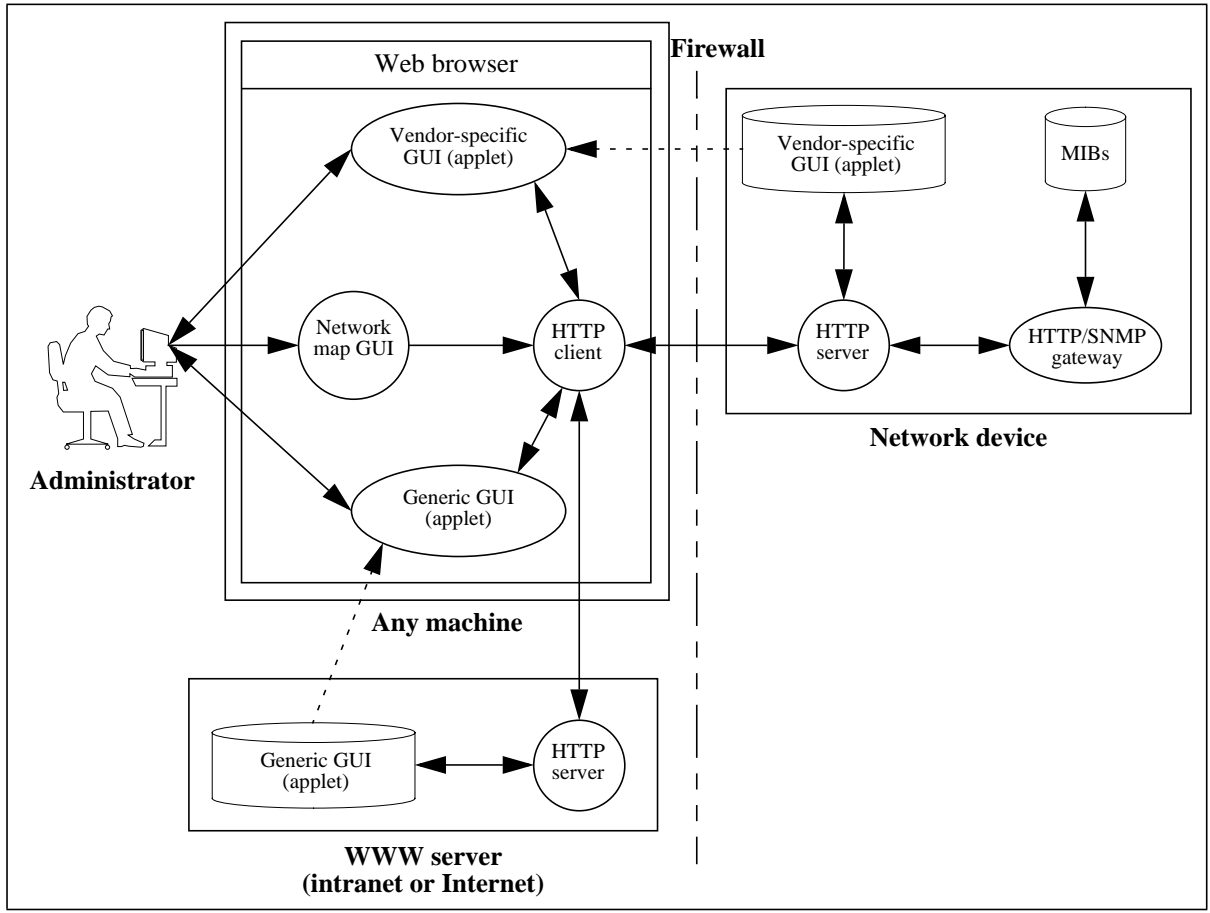

**Fig. 3.** Pull model: ad hoc management based on HTTP

In this section, we showed that SMEs that only rely on ad hoc management for their network equipment can save the cost of a network management platform by using Web technologies instead. Thus, depending on the pricing policy of network equipment vendors, the move from SNMP-based to Web-based network management can completely reshape the SMEs segment of this market. With SNMP-based ad hoc management, customers are charged for a management platform, per type-of-equipment add-ons, and per-device support for SNMP. With Web-based ad hoc management, they are only charged per-device support for HTTP. If the price of the latter is correctly defined by network equipment vendors, they can decrease the bill of customers while still increasing their revenue. Therefore, if network equipment vendors manage to keep the price of embedded HTTP servers low enough, they can eliminate the bottom-of-the-range network management platform market, and satisfy their customers by reducing their network management bill. In these conditions, Web-based network management is hard to resist for SMEs.

### 3.3. Regular Management

As we explained in section 3.2.1, many organizations do require regular management as well. For them, network management needs to be automated to a large extent, and includes data polling and event handling as well, which are not dealt with by the solutions we presented so far. Let us concentrate on data polling first. To begin with, let us suppose we have two separate management platforms, side by side: one for regular management, based on SNMP, and one for ad hoc management, based on Web technologies. We will describe, step by step, how to integrate them.

The first step is to integrate a Web browser in the network management platform[1], in order to have a unified interface for ad hoc and regular management. The network map GUI is loaded in the Web browser, and serves as the entry point to all subsequent interactions with the administrator or the operator (like for ad hoc management). But for ad hoc management, the network map GUI could be a static image, e.g. an HTML sensitive map. This time, we want the network map to be updated by the event correlator, with icons turning red, green, etc. To be able to change dynamically its GUI, the network map must be coded as an applet. In Fig. 4, we show that this applet is loaded from a WWW server, where it is stored in a *management software repository*. The simplest way to update this map is to open a socket between the applet and the *event correlator*, an object living in a Java application. But since we do not necessarily want to have a network map GUI[2], we added an intermediary, the *network map registry*, between the network map GUI(s) and the event correlator. Hence, we can have zero, one or several network map GUIs registered independently.

The applet security model requires that the applet be downloaded from the same machine where the Java application is running. So the WWW server at the bottom of Fig. 4 and the WWW server at the top right are actually the same machine. They are represented separately to make Fig. 4 more readable.

The second step is to use applets for all GUIs provided by the network management platform (see [13] for the definition of the role played by these GUIs in traditional network management platforms). These are the vendor-specific management applets, the generic management applets, the polling definition and scheduler applets, and the report definition and scheduler applets. The vendor-specific applets are loaded from network devices, as before. All other GUIs are independent from each other, and can therefore be decoupled in separate applets, as depicted in Fig. 4. To make this figure more readable, we assume here that the latter applets are loaded from the same WWW server, which is not necessarily the case.

The third step is to make the data repository independent of the network management platform. We assume in this paper that data is stored in a third-party RDBMS; but we may use another technology, such as a plain text file system, or an object-oriented database. The data repository is stored on a machine that we call the data server. This can be any machine, not necessarily the WWW server, or the machine where the Web browser is running. To store or retrieve data, we propose to use JDBC; other technologies are also possible, that implement some kind of glue between the applet and the data repository. In order to make Fig. 4 easier to read, we assume that we have a single data repository for polling and report definitions and schedules. Of course, this does not have to be. In the case where the data repository is an RDBMS, this is a very reasonable assumption to make.

The delivery of SNMP notifications and the handling of events will be investigated in next section. The problem is different from data polling, because notification transfers are initiated by the agent, not the manager. Here, we simply assume that we have some kind of glue code between event handlers, running in the traditional management platform, and our Java application, where the event correlator runs.

The fourth step is to implement data polling as a Java application using HTTP to collect data. Upon start-up, the polling engine retrieves all polling definitions and schedules from the RDBMS, and based on an internal clock, polls all agents via HTTP on a regular basis. The data retrieved for network monitoring is checked by the polling data interpreter, which may generate an event in case a problem is inferred. This event is then dispatched to the event correlator, which can in turn decide to update the network map with, say, a red icon.

The communication between the manager and the agent is based on HTTP in Fig. 4. The reason for this choice is to keep the agent simple: it only requires an HTTP server, a feature offered by more and more network devices today. Many other options exist, though, as we will see in section 4.

---

1. This feature is now offered by many commercial network management platforms.
2. Regular management can rely entirely on event handlers, when it runs in unattended mode [13]. In such a case, the administrator is warned (e.g., paged, emailed or telephoned) when a problem is deemed serious (see end of this section).

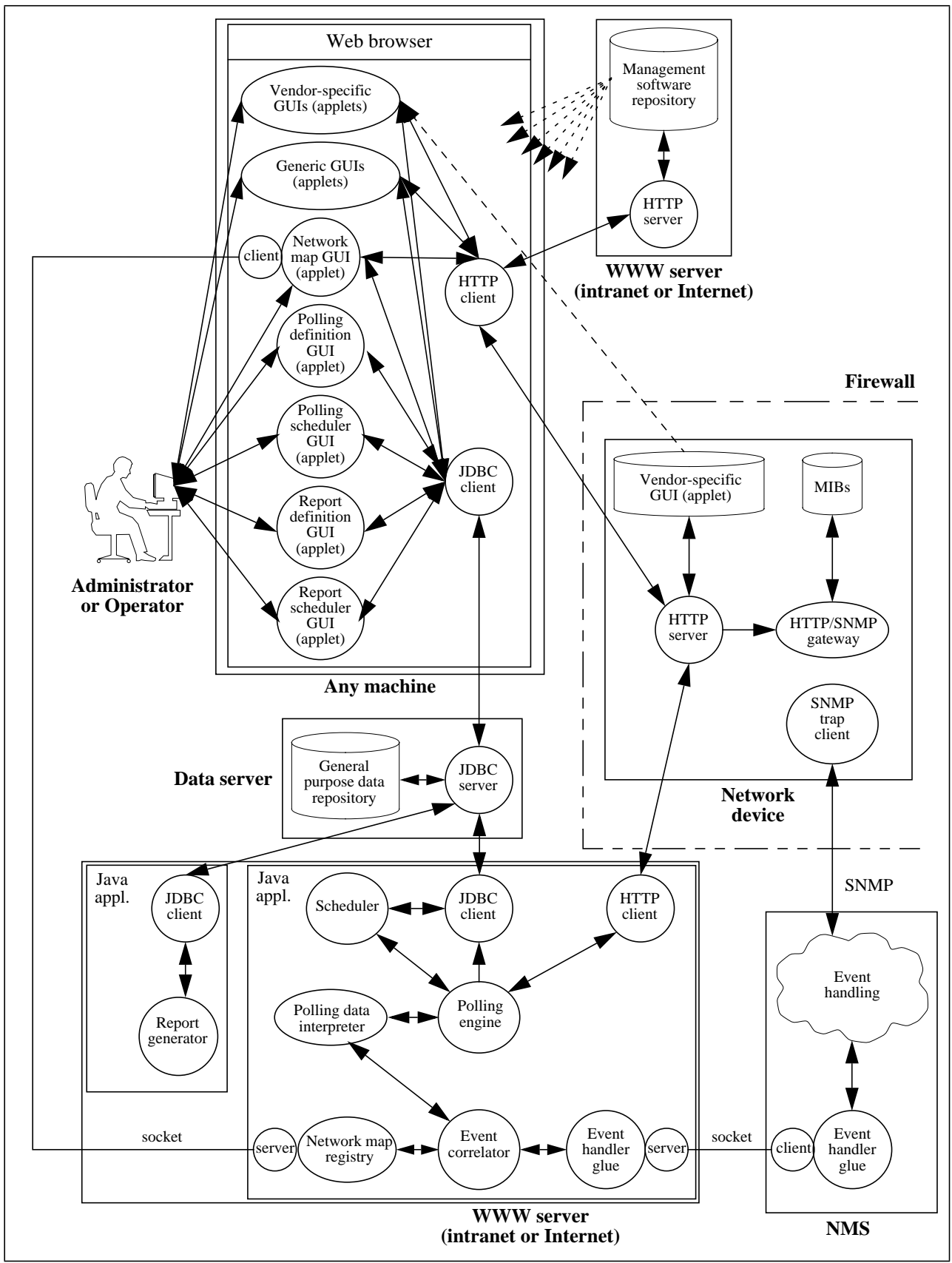

**Fig. 4.** Pull model: data polling based on HTTP

How does the polling engine know about modifications made via the polling definition applet or the polling schedule applet? A very inefficient solution would be for the polling engine to poll the data repository at regular time intervals. Alternatively, we could have one socket between the polling engine and the polling definition applet, and one socket between the polling engine and the polling scheduler applet. Otherwise, we could have an RMI client in the two applets, and an RMI server in the polling engine, so that each time a definition or schedule is altered (i.e., created, modified or deleted), the polling engine is informed via RMI. Yet another solution would consist in using an active database instead of a plain RDBMS, so that each time a polling definition or schedule is altered in the database, a trigger sends this information to the polling engine, via a socket or RMI.

The fifth and last step is to migrate reports generation to Web technologies. This can be achieved fairly simply with a Java application, accessing the data repository via JDBC. This data is stored by the polling application, and is accessed independently by the report generator. For the sake of simplicity, we pictured the two Java applications, report generation and data polling, as taking place on the same machine; but this does not have to be the case. The report definition applet and the report scheduler applet do not have to interact with the report generator directly, so there is no constraint put by the applet security model.

One case we have not discussed so far is when regular management is performed in unattended mode. Some companies rely on fully automated network management by running a network management platform performing regular management, without operators. When the event correlator detects what it deems to be a very serious problem, it contacts the administrator in an emergency mode adequate to the situation: siren, pager, email, etc. In this case, it is not necessary to permanently clog up a machine with a Web browser, since nobody looks at this browser. The event correlator only communicates with event handlers, which are appropriate for unattended mode, and no socket is connected to the network map registry.

At this stage, data collection and network monitoring rely entirely on Web technologies. They no longer require an expensive NMS, dedicated to network management. SNMP notifications delivery and event handlers are the only tasks still performed in the traditional way.

# 4. The Push Model

The push model, as we saw in section 3.1, is based on the publish/subscribe/distribute paradigm. It generalizes to network monitoring and data collection the way SNMP notifications are filtered and delivered today. The push model was recently put in the spot light by the large success encountered by push technologies[1] on the Web (e.g., Pointcast, Backweb, or Tibco's TIB software suite). Even though this model is well-known in software engineering, it has always been confined to SNMP notification delivery in IP network management: to the best of our knowledge, no management platform uses it for network monitoring or data collection today. Yet, we claim that its very design makes it better suited to regular management than the pull model. This section will now present how to use the push model in Web-based network management.

The chief advantage of using push technologies is to conserve network bandwidth, and move part of the CPU burden from managers to agents. Much of the network overhead caused by pull technologies is due to the fact that data collection and network monitoring are very repetitive: there is a lot of redundancy in what the manager keeps asking all the agents. For instance, in network monitoring, a common way to check if a machine is still alive is to request the same MIB variable, typically its `sysObjectID` (MIB-II), every few minutes. This scheme is very inefficient if the manager marshalls and sends this OID to all the agents, at every polling cycle, endlessly. Instead, with the push model, the manager contacts each agent once, subscribes to this OID once (push data definition), and specifies at what frequency (push frequency) the agent should send the value of this MIB variable (push data schedule); afterwards, there is no more traffic going from the manager to the agent: all subsequent traffic goes from the agent to the manager (except in the rare cases when the manager wishes to make a change in the list of OIDs it is interested in, or in the push frequency); the agent remembers what MIB variables the manager subscribed to, and how often it should send the value of these variables. The way the agent "remembers" this is by storing the push definitions and schedules on local storage; if this is EPROM, the agent can retrieve these definitions and schedules by itself after a reboot; if this is RAM, it needs to retrieve them from the manager, which stores them in the data server.

---

1. Despite their name, most push technologies actually follow the pull model at the implementation level. TIB Rendez-Vous is one of the rare exceptions.

The point of moving part of the CPU burden from managers to agents is to decrease the requirements put on NMSs in terms of CPU and memory. Network management platforms for large networks are often big Unix or Windows NT servers, which cost a fortune to buy and maintain. Agents, on the other hand, are more powerful than they used to be, and most agents can reasonably do a bit of processing locally (see the rationale behind Goldszmidt's Management by Delegation scheme [8], or Wellens and Auerbach's myth of the dumb agent [25]).

Another advantage of the push model is that it can be used in conjunction with multicasting to make network management more robust. When an administrator subscribes to some management data (that is, MIB variables or SNMP notifications), he/she tells the agent what manager it should send the data to. Instead of specifying a unicast IP address, he/she can specify a multicast address instead. For agents, sending data to a unicast or a multicast address is transparent; the only requirement is that they support IP multicasting, which modern implementations of the TCP/IP stack generally do. Management data is thus sent to multiple managers in parallel, which makes the network management system more robust; e.g., one manager can crash while another takes over transparently (so-called "hot standby"). When the pull model is used, we can also have the agent send management data to several managers; but in this case, all of the managers have to request the data independently, which increases significantly the network overhead, and also consumes more CPU cycles. Here, with the push model, standby managers can receive data passively until they are configured to replace the previous manager. Even if this is still a long way from fault-tolerant systems, this new feature can be very attractive to organizations whose network is of critical importance to the smooth running of their business.

Compared to the pull model, the push model introduces a new issue: synchronization. If the manager and the agent have internal clocks that do not synchronize regularly, they will probably drift apart. This is not a problem for network monitoring, since the administrator does not care, when it configures an agent to send a heart beat to its manager every 5 minutes, whether the manager receives it every 299 seconds or every 301. But it is a small problem for data collection, since the manager may receive too much data, or less than expected. To build reports, some data will have to be discarded arbitrarily, or some data interpolated. When the push model is used, it is therefore recommended to synchronize the clocks of network equipment on a regular basis. This may rely on the use of a protocol like NTP, or may require the manager to contact all agents every hour or so, and exchange a few synchronization packets. In both cases, the synchronization overhead is negligible compared to the network and CPU savings induced by going from pull to push.

Let us now investigate the engineering details of the push model. Section 4.1 will present the publish and subscribe phases, whereas section 4.2 will present different scenarios for the distribute phase.

### 4.1. Publish and Subscribe Phases

In the first phase, the network device (agent) publishes what MIBs it supports, and what SNMP notifications it can send to the manager. A simple way to implement this is to use applets, as depicted in Fig. 5. First, the user (administrator or operator) selects an agent on the network map applet, and loads from that agent a well-known HTML page[1] which lists all management applets stored on the agent. Every applet publishes one MIB (vendor-specific or generic) supported by the agent, except one, which publishes the SNMP notifications supported by this agent.

In the second phase, the administrator subscribes the manager to MIB variables and SNMP notifications. *MIB data subscription applets* allow him/her to select MIB variables as well as push frequencies. The push frequency can be specified at the MIB variable level: it need not be the same for all variables of a given MIB. The push frequency is equal to the polling frequency considered in section 3. Obviously, the *notification subscription applet* does not have to specify a push frequency, as notifications are inherently asynchronous. In fact, the notification subscription applet is simply a filter: it specifies what notifications the manager is interested in. Other notifications are discarded by the notification generator (see Fig. 6).

Instead of having several MIB data subscription applets (one per MIB), we could have one, single applet allowing the subscription to any MIB variable of any MIB. But just like people dislike using MIB browsers in traditional network management platforms because they are too basic, and prefer to use management GUIs customized for each MIB, people would not be happy if they had to subscribe to MIB data without visual aids customized for each MIB.

---

1. For example <URL:http://agent.domain/mgmt/mibs.html>.

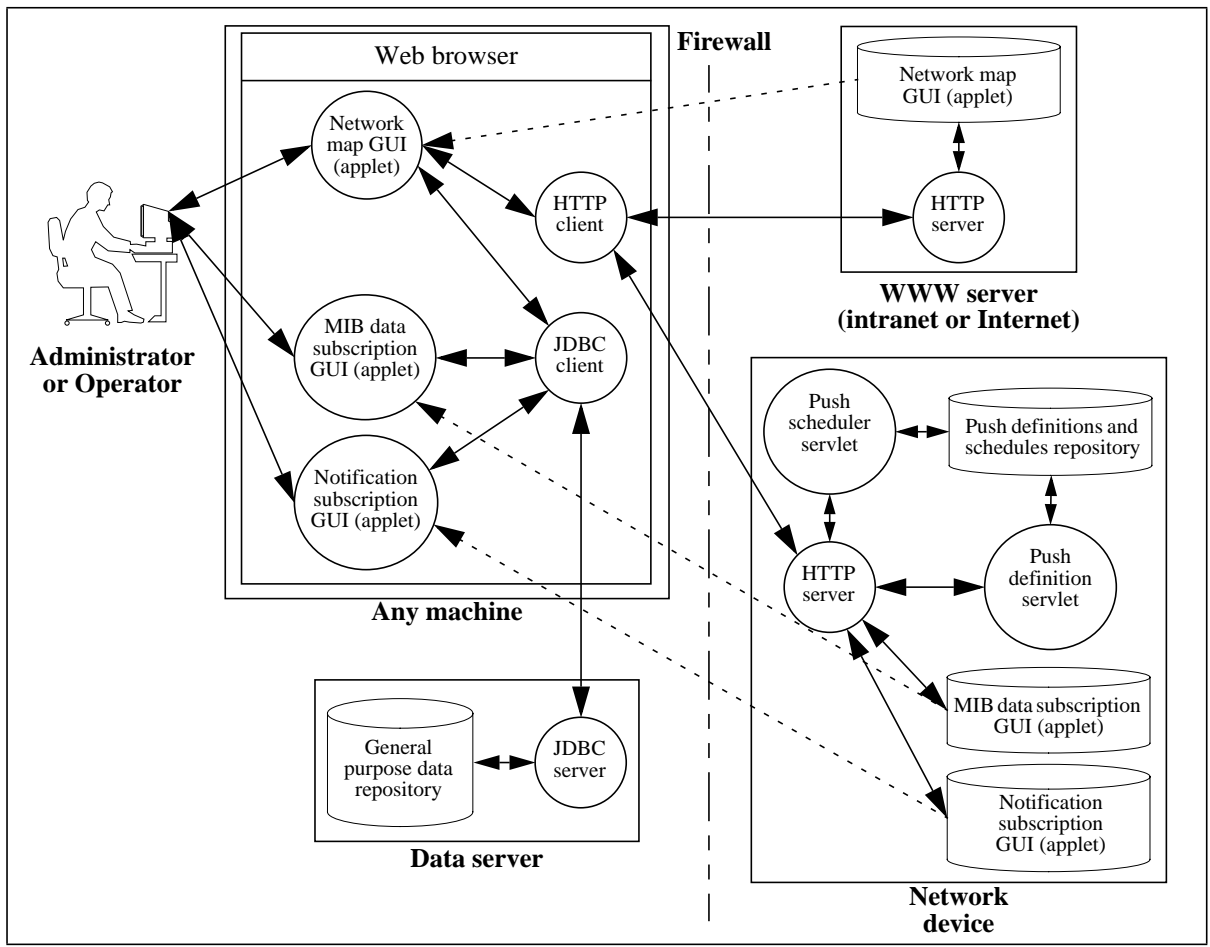

**Fig. 5.** Push model: publish and subscribe phases

The publish and subscribe phases are depicted in Fig. 5. The details of the MIB and notification subscriptions are stored in the data server. In case an agent loses all its push configuration data, this allows the manager to resend all the definitions and schedules for that agent in unattended mode: the administrator does not have to enter it all over again manually, via a GUI. The general purpose data repository of the data server includes (i) the definitions and schedules of the MIB data subscribed to by the manager, (ii) the definitions of the notifications subscribed to by the manager, and (iii) the network topology definition used by the network map applet to construct its GUI. In real life, these three logical data repositories may actually be stored into different databases, or a single database.

### 4.2. Distribute Phase

In the distribute phase, the case of data collection and network monitoring is only marginally different from the case of notification delivery and event handling. In order to facilitate the comparison with the pull model, we will concentrate on data collection and network monitoring in this paper, that is, how to replace data polling with push technologies. Notification delivery and event handling are presented in detail in [14]. But let us stress that the solutions we will describe below apply equally well to both cases: the communication issues between the agent and the manager are the same, only the Java applications running on the manager side are different.

The general purpose data repository depicted in Fig. 6 includes seven different repositories: the three listed in section 4.1, plus (iv) the event handler definitions repository, (v) the event handlers invocation log, (vi) the pushed data repository, and (vii) the pushed notifications repository. Like previously, in real life, all these logically different data repositories may actually reside in one or more databases.

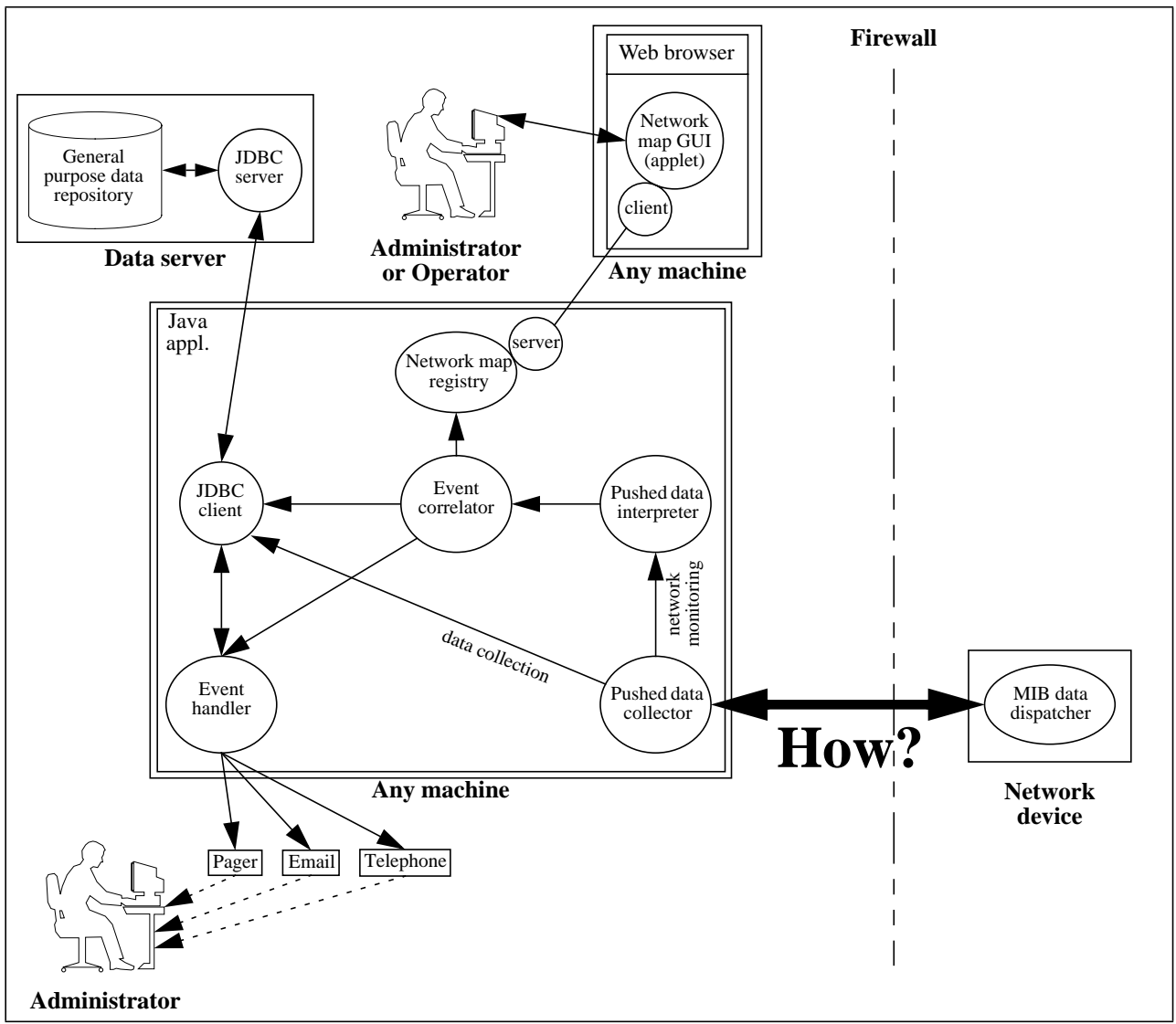

**Fig. 6.** Push model: distribute phase

Compared to Fig. 4 (pull model), we no longer have a polling engine; instead, we have a *pushed data collector*, that collects data related to network monitoring or data collection. This data is stored on the data server via JDBC. Since the execution speed of Java code is slow, it may be a good idea to increase the performance by storing data in bulk. Like the polling engine in the pull model, the pushed data collector sends data collected for network monitoring to the *pushed data interpreter*. If an abnormal condition is detected by the pushed data interpreter, e.g. a device no longer sends any data, an alarm is generated in the form of an event sent to the event correlator. The event correlator also receives events in the form of notifications (not shown here), and identifies the problem with the network. It can invoke an event handler, when an event is not masked by another, in which case the call to the event handler is logged in the data server.

The main difficulty when going from pull to push is that the data transfer is now initiated by the agent, instead of the manager, while the HTTP client remains on the manager side, and the HTTP server on the agent side (see Fig. 6). Somehow, the client and the server are on the wrong sides! We would like the server to initiate the communication, whereas communication is always initiated by the client in a client/server architecture. To address this issue, we have the choice between three communication technologies [21]: HTTP, sockets and RMI. For each of them, we will now propose ways of ensuring some kind of persistent connection[1] between the client and the server.

In this section, we will also address an issue already mentioned in sections 2.4 and 3.2.3: how to go across a firewall? This problem is not specific to the push model, and the engineering trade-offs that we will now describe apply equally well to the pull model (e.g., in Fig. 4, communication between the polling engine and the network device does not necessarily rely on HTTP).

---

1. The same connection would be used to deliver notifications [14].

**4.2.1. Sockets**

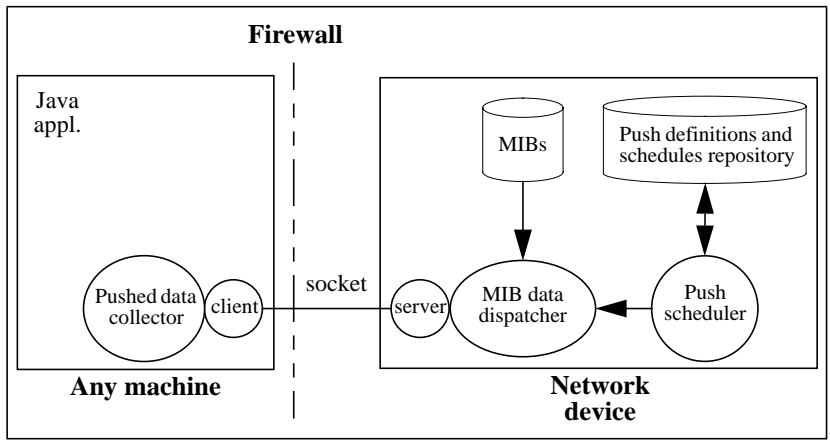

Fig. 7. Push model: distribution via sockets

The fact that sockets are bidirectional solves the problem of server-initiated communication: we can open a socket as usual, from the client to the server, and, later on, only use it to send MIB data from the socket server to the socket client. To ensure that this connection remains persistent, the *pushed data collector,* on the manager side, sets an infinite time-out value on the socket when it creates it. If the underlying TCP connection times out for whatever reason, it is the responsibility of the manager (i.e., the pushed data collector) to reconnect to the agent, by creating a new socket.

This socket-based solution presents a big advantage: simplicity. Sockets are very simple to program, especially in Java. But it also presents two potential drawbacks. First, if the underlying operating system of either the manager or the agent keeps timing out the connection (e.g., because the administrator has no control over the time-out value of the socket, and this time-out value happens to be lower than the push frequency), then this solution is clearly inappropriate. Not only do the repeated socket creations and time-outs cause network and CPU overhead, but even worse, we cannot take the risk to make notifications delivery depend on such a versatile type of persistent connection; there must be a way for the agent, not the manager, to create a new connection if the previous times out. Second, if we need to go across a firewall between the manager and the agent, there is a potential issue with sockets. As we saw in section 2.4, most firewalls filter out UDP, and let only a few TCP ports go through. So whether we use TCP or UDP sockets, firewalls will generally not let sockets go through by default. Thus, in order for this socket-based solution to work, the firewall system needs to be modified. This may not be a problem for large organizations, as we mentioned in section 2.4, because they either have in-house expertise in firewalls to set up UDP relays or change TCP filtering rules, or they can afford expensive external consultants to do the job. But it may well be a problem for SMEs, who generally lack this kind of expertise, and for whom expensive external consultants may not be an option.

**4.2.2. RMI**

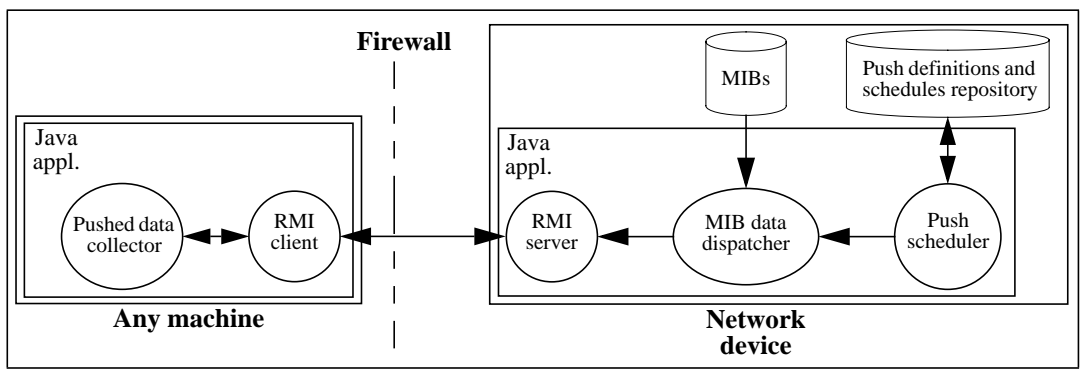

Fig. 8. Push model: distribution via RMI

Like sockets, RMI offers a bidirectional association: once an RMI client has bound to an RMI server, both of them can send data to the other. RMI is an elegant solution in terms of design, because it gives a fully object-oriented view of network management. It offers semantics to the network management application designer that are higher than mere MIB variables, and makes it easier to design complex applications [12].

But RMI also presents several drawbacks. First, it requires a full JVM to be embedded in all agents (as opposed to a light-weight JVM such as the one included in the EmbeddedJava platform). Very few network devices offer this feature today, primarily because Java chips have not encountered the success expected by the Java camp. Moreover, many devices will not have a full JVM for some time, because of the large footprint of this software in bottom-of-the-range devices that are very price sensitive (e.g., printer servers). Second, current RMI implementations are very slow, and use a lot of resources in terms of memory and CPU; therefore, RMI-based network management is not scalable. This may improve in future implementations, but the fact that other distributed object-oriented platforms such as CORBA or DCOM suffer from the same problems incites us to believe that object-oriented technologies such as RMI will remain, at best, a niche market in network management for the years to come. Third, RMI communication is actually based on sockets, which are transparent to applications; so once again, we face a problem with firewalls. In fact, things are even worse with RMI than in the previous case, because we no longer control what ports are used by sockets. RMI sockets are transparent to the application, so even if RMI servers run on a well-known port (`1099/tcp` or `1099/udp` [9]), RMI clients may bind to any port (in the previous case, the administrator could control what ports were used on the client and server sides). As a result, in order to use RMI across a firewall, one must add RMI-specific software to one's firewall system; and RMI relays are not supported by all firewall systems today (they may be in the future, if RMI proves successful over time).

For all these reasons, we cannot reasonably expect a large proportion of network equipment to support RMI in the near future. For the time being, sockets appear to be a better communication style for network management.

### 4.2.3. HTTP

HTTP does not share the property we exploited for sockets and RMI. It is not possible for the HTTP server to initiate a data transfer via a pre-existing persistent connection. All HTTP 1.1 methods rely on a request/response protocol, so a server cannot send a response without having received a request beforehand. It is not possible to work around this by having the HTTP server send an infinitely large number of responses to a single request from an HTTP client: a response may be fragmented, but the use of the 100 (Continue) status is limited [7]. In this respect, SNMP and HTTP are different. Both are based on the client/server model of communication; but SNMP is a request/response protocol for all operations (`get`, `set`, `inform`...) except one, `snmpv2-trap`, which relies on a one-way asynchronous transfer protocol; HTTP, conversely, is a request/response protocol for all methods (`get`, `post`, `head`...).

In order to allow HTTP-based communication between the manager and the agent, we therefore have to rely on a different engineering solution. We propose to add an HTTP 1.1 client on the agent, and an HTTP 1.1 server on the manager, so that client/server communication goes the normal way.

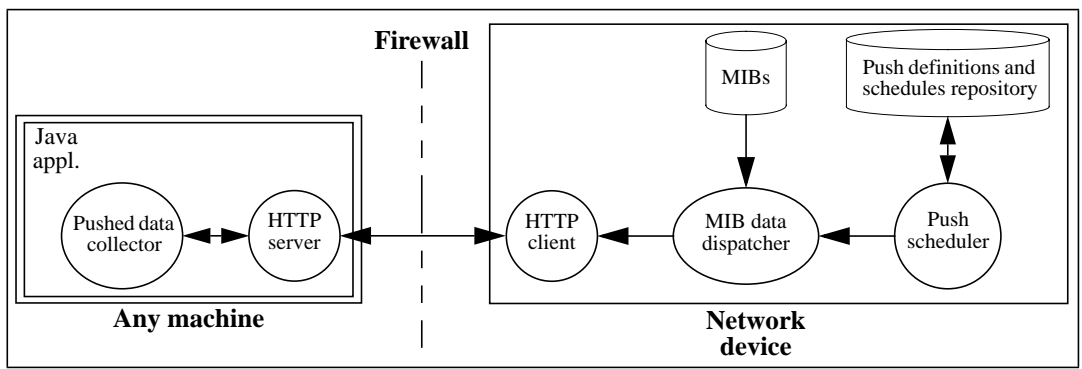

**Fig. 9.** Push model: distribution via HTTP

This solution presents several advantages. First, it does not rely on non-intuitive designs like in sections 4.2.1 and 4.2.2: the client is on the agent side, and the server on the manager side. Second, the agent can reconnect immediately in case the persistent connection times out: it does not have to count on the manager to do that; this improves the robustness, and avoids time windows when the agent wants to send data to the manager, but the manager has not reconnected to the agent yet. Third, no change at all is required on the firewall system, if the management application runs on the external Web server of the organization; if it runs on a different machine, a minor change in the setup of the firewall system is needed. The main drawback of this solution is that it requires an HTTP server to be included in the Java application running on the manager side. This makes a large program (the network management application) even larger, more difficult to debug, slower to execute, and induces a larger footprint on the machine where it is running.

# 5. Collapsed network management platform

If we combine Fig. 4, Fig. 5 and Fig. 6, we realize that we have completely dismantled the network management platform. We no longer need an expensive management station, dedicated to network management, and running expensive platform-specific software. Instead, we can download applets in Web browsers running anywhere, we can run Java applications on any machine fitted with a JVM, we can use existent RDBMSs to store management data repositories, and we can use internal WWW servers which are already ubiquitous. We call this the *collapsed network management platform*, by analogy with the concept of *collapsed backbone*.

# 6. Conclusion

Web-based network management is not just a fashion. We demonstrated in this paper that it is based on sound technical grounds, and offers cost-effective alternatives to the traditional IP network management platforms used today in the industry. We believe it is promised to a bright future, as it yields significant savings to almost all actors in this market. First, it enables customers to save the cost of expensive network management platforms, and helps capitalize on existing investments such as RDBMSs. Second, it allows network equipment vendors to cut drastically the development costs of their vendor-specific management GUIs; they no longer need to port add-ons to many management platforms running on different operating systems; now, they can develop a single applet that can be downloaded and executed by any Web browser running on any machine. Third, it gives a competitive edge to vendors moving toward Web technologies, by allowing them to reduce the time-to-market of their proprietary GUIs down to zero. Finally, it puts start-up companies in fair competition with larger, well-established companies, by giving them access to integrated network management, despite their small market shares. In fact, only network management platform vendors lose out with the Web. They cannot force customers to use proprietary databases anymore, and they cannot charge high prices for the few Java applications and generic GUI applets they sell.

In this paper, we presented two design approaches based on Web technologies: the pull model, well suited to ad hoc management, and the push model, well adapted to regular management. Engineering solutions were presented for data collection and network monitoring for both models. A companion paper [14] describes how to perform notification delivery and event handling with the push model. To implement the push model, the changes required in managed devices are limited when data distribution relies on HTTP or sockets. For the pull model, they are even more limited: we only need an HTTP server and an HTTP-to-SNMP gateway. None of these models mandates that a full JVM be embedded in all agents: top-of-the-range network equipment can optionally use RMI.

We will shortly begin implementing a prototype of a Java-based network management platform, as proof of the concepts presented herein. This platform will support both the pull and the push models. The test equipment to be managed will be provided by Lightning, a router vendor located in Lausanne, Switzerland. We will add a push scheduler and a MIB data dispatcher to their management software, which already includes an HTTP server, and will develop a vendor-specific applet to support their proprietary MIB. We hope to report at the conference, and integrate in the final paper, results and experience gained with this prototype.

# Acknowledgments

This research was partially funded by the Swiss National Science Foundation (FNRS) under grant SPP-ICS 5003-45311. The author would like to thank J. Schönwälder for discussions on the efficiency of BER encoding and SNMP, G. Madhusudan for discussions on communication in distributed Java applications, and H. Cogliati for proofreading this paper.

# Acronyms

| API | Application Programming Interface | NMS | Network Management Station |
| --- | --- | --- | --- |
| ASN.1 | Abstract Syntax Notation 1 | NTP | Network Time Protocol |
| BER | Basic Encoding Rules | OID | Object IDentifier |
| CORBA | Common Object Request Broker Architecture | PC | Personal Computer |
| CPU | Central Processing Unit | PER | Packed Encoding Rules |
| DBMS | DataBase Management System | RAM | Random Access Memory |
| DCOM | Distributed Common Object Model | RFC | Request For Comment |
| EPROM | Electrically erasable Programmable Read-Only Memory | RMI | Remote Method Invocation |
| GIF | Graphics Interchange Format | SME | Small to Medium-sized Enterprise |
| GUI | Graphical User Interface | SMI | Structure of Management Information |
| HTML | HyperText Markup Language | SNMP | Simple Network Management Protocol |
| HTTP | HyperText Transfer Protocol | SSL | Secure Sockets Layer |
| IETF | Internet Engineering Task Force | TCP | Transmission Control Protocol |
| IP | Internet Protocol | TLS | Transport Layer Security |
| JDBC | Java DataBase Connectivity | UDP | User Datagram Protocol |
| JDK | Java Development Kit | UPS | Uninterruptible Power Supply |
| JIT | Just In Time | URL | Uniform Resource Locator |
| JMAPI | Java Management Application Programming Interface | VPN | Virtual Private Network |
| JVM | Java Virtual Machine | W3C | World-Wide Web Consortium |
| LAN | Local-Area Network | WAN | Wide-Area Network |
| MIB | Management Information Base | WBEM | Web-Based Enterprise Management |
| MTBF | Mean Time Between Failures | WWW | World-Wide Web |

## Biography

J.P. Martin-Flatin is currently preparing for a Ph.D. thesis at EPFL. From 1990 to 1996, he was with the European Centre for Medium-Range Weather Forecasts in Reading, England, where he worked in network and systems management, security, Web management and software engineering. From 1988 to 1990, he worked on the Geographic Information System of a large city in France. In 1986, he received an M.Sc. in a mix of EE and ME from ECAM, Lyon, France. His main research interest is in distributed network management. He is a member of the IEEE and the ACM.